\documentclass{emulateapj}
\usepackage{amssymb}   
\usepackage{colordvi}   

\shorttitle{IMF--metallicity relation} \shortauthors{I. Mart\'in-Navarro et al. }

\def\gsim{ \lower .75ex \hbox{$\sim$} \llap{\raise .27ex \hbox{$>$}} }
\def\lsim{ \lower .75ex \hbox{$\sim$} \llap{\raise .27ex \hbox{$<$}} }
\newcommand{\msun}{\hbox{M$_{\odot}$}}

\begin{document}
\title{IMF -- metallicity: a tight local relation revealed by the CALIFA survey}

\author{Ignacio Mart\'in-Navarro$^{1,2}$, Alexandre Vazdekis$^{1,2}$, Francesco La Barbera$^{3}$, Jes\'us Falc\'on-Barroso$^{1,2}$, \\
 Mariya Lyubenova$^{4}$, Glenn van de Ven$^{5}$, Ignacio Ferreras$^{6}$, S.F. S\'anchez$^{7}$, S.C. Trager$^{4}$,  R. Garc\'ia-Benito$^{8}$, \\
 D. Mast$^{9}$, M.A. Mendoza$^{8}$, P. S\'anchez-Bl\'azquez$^{10}$, R. Gonz\'alez Delgado$^{8}$, C.J. Walcher$^{11}$ \\ 
 and the CALIFA team}
\affil{$^{1}$Instituto de Astrof\'{\i}sica de Canarias,c/ V\'{\i}a L\'actea s/n, E38205 - La Laguna, Tenerife, Spain}
\affil{$^{2}$Departamento de Astrof\'isica, Universidad de La Laguna, E-38205 La Laguna, Tenerife, Spain} 
\affil{$^{3}$INAF - Osservatorio Astronomico di Capodimonte, Napoli, Italy}
\affil{$^{4}$Kapteyn Astronomical Institute, University of Groningen, Postbus 800, 9700 AV Groningen, The Netherlands} 
\affil{$^{5}$Max Planck Institute for Astronomy, K\"onigstuhl 17, 69117, Heidelberg, Germany} 
\affil{$^{6}$Mullard Space Science Laboratory, University College London, Holmbury St Mary, Dorking, Surrey RH5 6NT}
\affil{$^{7}$Instituto de Astronom\'ia, Universidad Nacional Auton\'oma de M\'exico, A.P. 70-264, 04510, M\'exico, D.F.}
\affil{$^{8}$Instituto de Astrof\'{\i}sica de Andaluc\'{\i}a (CSIC), Glorieta de la Astronom\'\i a s/n, Aptdo. 3004, E18080-Granada, Spain} 
\affil{$^{9}$Instituto de Cosmologia, Relatividade e Astrof\'{i}sica - ICRA, Centro Brasileiro de Pesquisas F\'{i}sicas, Rua Dr.Xavier Sigaud 150, CEP 22290-180, Rio de Janeiro, RJ, Brazil}
\affil{$^{10}$Departamento de F\'{\i}sica Te\'orica, Universidad Aut\'onoma de Madrid, Cantoblanco, E28049, Spain}
\affil{$^{11}$Leibniz-Institut f\"ur Astrophysik Potsdam (AIP), An der Sternwarte 16, D-14482 Potsdam, Germany}
\email{email: imartin@iac.es}

\begin{abstract}

Variations in the stellar initial mass function (IMF) have been invoked to explain the spectroscopic and dynamical properties of early-type galaxies. However, no observations have yet been able to disentangle the physical driver. We analyse here a sample of 24 early-type galaxies drawn from the CALIFA survey, deriving in a homogeneous way their stellar population and kinematic properties. We find that the local IMF is tightly related to the local metallicity, becoming more bottom-heavy towards metal-rich populations. Our result, combined with the galaxy mass--metallicity relation, naturally explains previous claims of a galaxy mass--IMF relation, derived from non-IFU spectra. If we assume that -- within the star formation environment of early-type galaxies -- metallicity is the main driver of IMF variations,  a significant revision of the interpretation of galaxy evolution observables is necessary.

\end{abstract}


\section{Introduction}

The stellar initial mass function (IMF) determines the ratio between dwarf and giant stars, and therefore, ultimately regulates the chemical enrichment and the stellar feedback. Moreover, it is the \emph{Rosetta stone} in our understanding of unresolved stellar systems, relating the properties of resolved nearby stars to the integrated light of more distant objects.

Over the last years, there is growing evidence of a variable IMF, as opposed to the common assumption that the IMF of the Milky Way is universal \citep{kroupa,bastian}. These claims come from a wide variety of approaches, including stellar population analysis \citep{cenarro,vandokkum,spiniello12,ferreras}, gravitational lensing \citep{treu,thomas11} and dynamical models \citep{cappellari}. 

Assuming that the answer to all of these observational results is indeed a non-universal IMF, two main parameters have been proposed. On the one hand, the integrated stellar velocity dispersion is found to correlate with the IMF slope as inferred from both dynamical models \citep[e.g.][]{cappellari} and stellar population analysis \citep[e.g.][]{labarbera}. On the other hand, \citet{cvd12} claimed a stronger correlation when comparing the [Mg/Fe] pattern to the IMF slope. Nevertheless, these trends have to be carefully interpreted because they are based on the spatially-integrated light of early-type galaxies (ETGs). Thus, the inferred properties are luminosity weighted quantities, convolved by the radial light distribution, which depends itself on the stellar velocity dispersion \citep{Graham}, and is highly peaked towards the centre.

We showed in \citet[][see also \citealt{pasto}]{mn14,mn14b} that the IMF is a local property, not only varying among but also within galaxies. The lack of a large statistical sample, however, limited further interpretations of our data, and the question about which is the main driver of IMF variations remains open. In this Letter, we use the 2D spatially-resolved IFU data from the CALIFA survey \citep{califa} to address this problem by analyzing the radial IMF profiles of a sample of 24 ETGs.

\begin{figure*}
\begin{center}
\includegraphics[width=14.5cm]{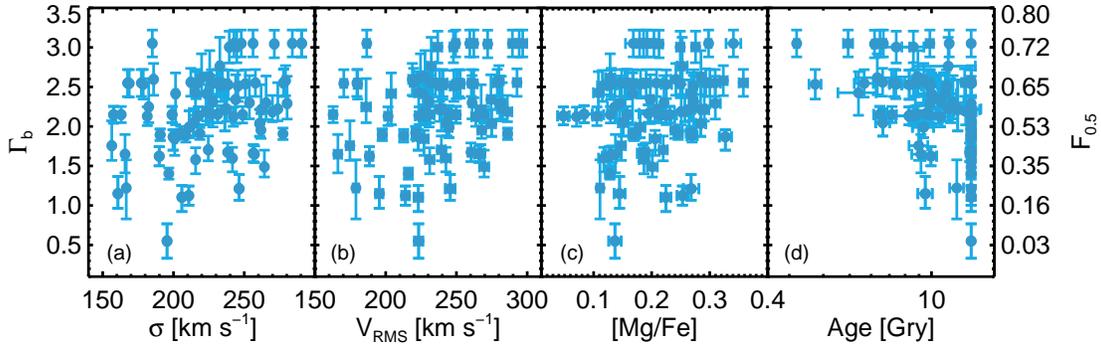}
\end{center}
\caption{The best-fitting IMF slope $\Gamma_\mathrm{b}$ is compared to the local $\sigma$ (a), $V_\mathrm{rms}$ (b), [Mg/Fe] (c) and age (b). Neither the kinematics properties nor the [Mg/Fe] or the age follow the measured IMF variations ($\rho_\mathrm{\sigma}=0.35$, $\rho_{V_\mathrm{rms}}=0.30$, $\rho_\mathrm{[Mg/Fe]}=0.21$, $\rho_\mathrm{age}=-0.50$, with $\rho$ being the Spearman correlation coefficient). The right vertical axis represents the IMF slope in terms of F$_{0.5}$, defined as the fraction (with respect to the total mass) of stars with masses below 0.5 M$_\sun$.}
\label{fig:all}
\end{figure*}

\section{Data}

We selected the sub-sample of 24 ETGs with redshift 0.018\,$<$\,$z$\,$<$\,0.030 among the observed CALIFA galaxies \citep[for sample properties see][]{mother}. In the local Universe, the strong IMF sensitive index TiO$_2$ is potentially affected by a telluric absorption feature at $\sim$6280\AA. The selected redshift window maximizes the number of ETGs unaffected by this feature. Technical details of the data are fully described in \citet{califa,dr2}. The central velocity dispersion in our galaxy sample ranges from $\sim$160 to $\sim$310 km\,s$^{-1}$, although the bulk of them (21 galaxies) have velocity dispersions greater than 200 km\,s$^{-1}$. The mean stellar mass in our sample is $ \overline{M}_\star=10^{11.54} M_\sun$ according to \citet{Rosa}.

We derived the line-of-sight mean stellar velocity (V) and velocity dispersion ($\sigma$) following Falc\'on-Barroso et al. (2015, \emph{in prep.}) Using these measurements, each IFU spaxel was then corrected to the rest-frame. Finally, we radially binned each galaxy using elliptical apertures. The size of the apertures was set to reach a signal-to-noise of, at least, 125 per \AA \ at $\lambda\lambda$~=~6000, 6200\AA. All the quantities discussed in this Letter are averaged spaxel-values over these elliptical annuli. \looseness-1

\section{Analysis}\label{sec:ana}

\subsection{Stellar populations}
For this work, we made use of the extended MILES (MIUSCAT) stellar population models \citep{miles,miuscat}. We assume a bimodal, low-mass tapered, IMF, whose only free parameter, $\Gamma_\mathrm{b}$, is the slope of the high-mass end (above 0.6\msun) of the distribution. This parametrization, first introduced by \citet{vazdekis96}, generalizes the Kroupa IMF, which is recovered for $\Gamma_\mathrm{b}=1.35$. Note that the current version of the extended MILES models cover from $\Gamma_\mathrm{b}=0.3$ to $\Gamma_\mathrm{b}=3.3$, which can lead to some saturation for high IMF values.

Given the CALIFA wavelength range ($\lambda \lambda$~3700,\,7500\,\AA), we focused on five prominent spectral indices: H$_{\beta_\mathrm{O}}$, [MgFe]$^\prime$, Mg2Fe \citep{bc03}, NaD, TiO$_{2_{\mathrm{CALIFA}}}$ and TiO$_1$. The last three are IMF-sensitive features, whereas [MgFe]$^\prime$ and Mg2Fe depend on metallicity and  H$_{\beta_\mathrm{O}}$ on age. The TiO$_{2_{\mathrm{CALIFA}}}$ index follows the standard TiO$_2$ definition \citep{trager} but with a narrower blue pseudo-continuum ($\lambda\lambda$~6060.625,~6080.625\,\AA) to avoid any telluric contamination. The correction of H$_{\beta_\mathrm{O}}$ from nebular emission was done in the same way as in \citet{labarbera}. For each spectrum (i.e. each galaxy and radial bin), we inferred stellar population properties with three different approaches:

i) Following \citet{labarbera}, we minimized: 
\begin{equation}
\chi^2(\Gamma_\mathrm{b},\mathrm{age},{\rm[M/H]})= 
  \sum_i\left[\frac{(EW_i-\Delta_{\alpha,i})-EW_{M,i}}{\sigma_{EW_i}}\right]^2,
  \label{eq:method}
\end{equation} 
\noindent where our index measurements ($EW_i$), after being corrected for non-solar abundances ($\Delta_{\alpha,i}$), are compared to the model predictions ($EW_{M,i}$). The fitting process was repeated by considering different combinations of IMF sensitive indices, leading to consistent results.

ii) We implemented an iterative fitting scheme where age and metallicity were calculated first using an index-index (H$_{\beta_\mathrm{O}}$-[MgFe]$^\prime$) diagram, assuming a standard IMF. The age estimate coming from this first step was then used to derive the metallicity and the IMF in a second index-index (TiO$_{2_{\mathrm{CALIFA}}}$[MgFe]$^\prime$) diagram. These two steps were repeated, re-deriving age and metallicity from the H$_{\beta_\mathrm{O}}$-[MgFe]$^\prime$ diagram with an updated IMF, until the solution converged.

iii) Alternatively, we also fit our data following Eq.~\ref{eq:method} but neglecting individual [$\alpha$/Fe] corrections. 

\begin{figure*}
\begin{center}
\includegraphics[width=10.5cm]{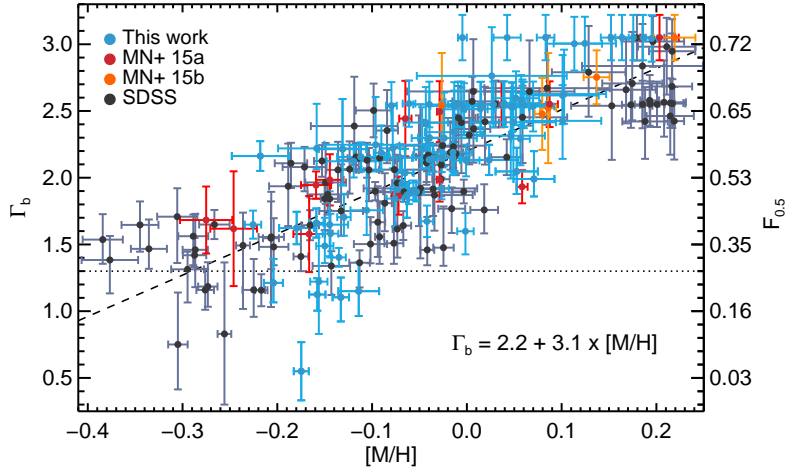}
\end{center}
\caption{IMF--metallicity relation obtained from CALIFA local measurements (blue). We also show the local IMF and metallicities measurements derived by \citet{mn14,mn14b} (red, orange) for three of nearby ETGs, as well as global SDSS measurements (black). We found it to be the strongest correlation ($\rho_\mathrm{[M/H]}=0.82$). As in Fig.~\ref{fig:all}, the right vertical axis indicates the F$_{0.5}$ ratio. For reference, the standard Kroupa IMF value is shown as a horizontal dotted line. Dashed line correspond to the best-fitting linear relation to all the datasets.}
\label{fig:met}
\end{figure*}

The three approaches show a good agreement, with small differences only for low IMF-slope values, where IMF effects can be mimicked by changes in other stellar population parameters. For simplicity, we will refer through this Letter to the best-fitting values derived from the simultaneous fit (Eq.~\ref{eq:method}) of the H$_{\beta_\mathrm{O}}$, [MgFe]$^\prime$ and TiO$_{2_{\mathrm{CALIFA}}}$ indices. Note that the latter, compared to the TiO$_1$, exhibits a milder dependence on both abundance ratios and total metallicity \citep{labarbera}. However, the absolute TiO$_2$ sensitivity to these parameters depends on the adopted stellar population synthesis model \citep{bc03,thomas11}. Moreover, TiO$_2$ is less sensitive to variations in the temperature scale of giant stars. \citet{spiniello15} showed that, after accounting for metallicity, no variation in the effective temperature is needed to fit the strength of gravity-sensitive features in massive ETGs, and that a similar result would be recovered with other stellar population models if the same IMF parameterization were used. Although our fitting scheme does not account for the residual impact of non-solar abundances on the temperature scale of the (solar-scaled) isochrones, the [$\alpha$/Fe] effect on the isochrones is significantly milder than that on the stellar atmospheres \citep{alpha}. The latter is corrected in our approach ($\Delta_{\alpha,i}$ in Eq.~\ref{eq:method}).

In addition to the IMF, age and metalliciy, the [Mg/Fe] of each radial bin was derived by means of the [Z$_\mathrm{Mg}$/Z$_\mathrm{Fe}$] proxy \citep{labarbera}, i.e., using the metallicity difference between two index-index diagrams, where H$_{\beta_\mathrm{O}}$ is plotted against a Mg and Fe metallicity indicator, respectively.

\subsection{Stellar kinematics}
To understand the IMF variations, we also compared our best-fitting IMF values to two kinematics parameters: the local $\sigma$ and the local $V_\mathrm{rms}$ defined as $V_\mathrm{rms}\equiv\sqrt{V^2+\sigma^2}$ . Whereas $\sigma$ has been claimed to be the main driver of the IMF variations \citep{treu,ferreras}, in spatially resolved studies $V_\mathrm{rms}$ accounts for both random and ordered motions.

\section{Results}

In Fig.~\ref{fig:all} we present the correlation between the best-fitting IMF slope, $\Gamma_\mathrm{b}$, and the local values of [Mg/Fe], age, $\sigma$ and $V_\mathrm{rms}$. None of them show a tight correlation with $\Gamma_\mathrm{b}$. The mild relation between age and $\Gamma_\mathrm{b}$ can be understood either as a residual degeneracy between both parameters or as a consequence of the IMF-metallicity relation, since young stars within massive ETGs are likely formed from metal-enriched material.

\subsection{The IMF--metallicity relation}
Among all the explored relations, the IMF slope--local metallicity relation emerges as the most fundamental. This is shown in Fig.~\ref{fig:met}, where the local $\Gamma_\mathrm{b}$--[M/H] relation derived from the CALIFA survey (blue) is combined with local IMF estimates, obtained at different galactocentric distances, for three nearby ETGs by \citet{mn14,mn14b} (red and orange symbols). In addition, we also show (black) the best-fitting IMF and metallicity inferred from SDSS stacked spectra. To construct these spectra, we followed \citet{labarbera}, but binning according to both $\sigma$ and [M/H] of the individual galaxies. The broad wavelength range of the SDSS data set allows us to infer the IMF, not only using those features within the CALIFA spectral range, but also prominent near-IR IMF sensitive features such as the NaI\,8189 and the CaII triplet \citep[see \S4.1 in][for a detailed description of gravity-sensitive features in SDSS spectra]{labarbera}.

The fact that the three datasets included, although based on different sets of line-strengths, lie on the same relation, supports a tight connection between IMF slope and metallicity, regardless of the details in the determination of the stellar population parameters. Moreover, the agreement between integrated measurements from the SDSS spectra and spatially-resolved values, suggests that the mechanism behind the local IMF variations ultimately shapes the global galaxy mass--IMF relation. 

A linear fit to all the measurements shown in Fig.~\ref{fig:met} leads to the following relation between IMF slope and metallicity in ETGs
\begin{equation}
\Gamma_\mathrm{b}=2.2(\pm0.1)+3.1(\pm0.5)\times\mathrm{[M/H]}
\label{eq:fit}
\end{equation} 
Since IMF-sensitive features ultimately trace the dwarf-to-giant ratio F$_{0.5}$, as defined in \citet{labarbera}, the above equation can be expressed in terms of a single power law IMF as
\begin{equation}
\Gamma=1.50(\pm0.05)+2.1(\pm0.2)\times\mathrm{[M/H]}
\end{equation} 

Apart from the measurement errors, the scatter in the relation comes from two sources: the IMF--[$\alpha$/Fe] degeneracy when fitting gravity-sensitive features around the Kroupa-like IMF regime \citep{labarbera} and the dependence of the IMF on the minimized set of indices\footnote{Uncertainties in Eq.~2-3 account for this effect, by repeating the fit using only CALIFA data.} \citep{spiniello2013}. In this sense, the fact that the TiO$_2$-based CALIFA measurements show a steeper IMF-metallicity trend is consistent with a stronger metallicity dependence of this index than predicted in the MILES models. On the other hand, the consistency among different data-sets ($\rho=0.82$ when CALIFA, SDSS and \citet{mn14,mn14b} are considered) points to a genuine IMF-[M/H] trend, as shown in Fig.~\ref{fig:met}.

\subsection{The [MgFe]$^\prime$--TiO$_{2_{\mathrm{CALIFA}}}$ empirical relation}

\begin{figure}
\begin{center}
\includegraphics[width=8.5cm]{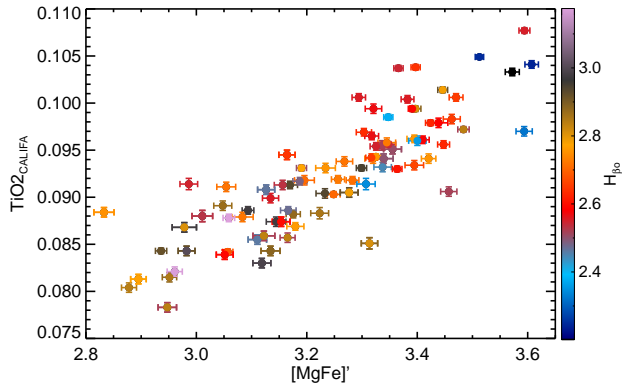}
\end{center}
\caption{Empirical relation between the metallicity-sensitive [MgFe]$^\prime$ and the IMF-sensitive TiO$_{2_{\mathrm{CALIFA}}}$ features. Index measurements (at a resolution of 200~km\,s$^{-1}$) are color-coded by their H$_{\beta_\mathrm{O}}$ value, as an age proxy. An IMF--metallicity relation is needed to explain the observed trend, since the TiO$_{2_{\mathrm{CALIFA}}}$ weakly depends on the total metallicity and [MgFe]$^\prime$ is almost independent of the IMF.}
\label{fig:mgtio}
\end{figure}

To strengthen the validity of our result, we adopted an empirical approach. In Fig.~\ref{fig:mgtio} we compare the [MgFe]$^\prime$ to the TiO$_{2_{\mathrm{CALIFA}}}$ individual measurements for all radial bins in our sample, at a common 200~km\,s$^{-1}$ resolution. Each point is color-coded by its H$_{\beta_\mathrm{O}}$ value, representing an age segregation. The [MgFe]$^\prime$ index is independent of the IMF, depending almost entirely on the total metallicity. On the contrary, the TiO$_{2_{\mathrm{CALIFA}}}$ is a strong IMF indicator, that increases with age \citep{labarbera}. Since MILES stellar population models predict the TiO$_{2_{\mathrm{CALIFA}}}$ feature to be, in the explored metallicity and age regime, almost independent of total metallicity, the strong (correlation coefficient $\rho=0.86$) correlation shown in Fig.~\ref{fig:mgtio} can be interpreted as a metallicity--IMF trend. A similar relation was found by \citet{trager} while comparing the Mg$_2$ and the TiO$_2$ indices. Note that if the TiO$_2$ sensitivity to total metallicity is larger than predicted by MILES models, a certain correlation between these two indices would also be expected.

\section{Discussion}

ETGs are characterized by old stellar populations and thus, only stars  with masses $M \lesssim1M_\sun$ remain alive. Therefore, the analysis of gravity-sensitive features in the integrated light of unresolved ETGs can only constraint the the dwarf-to-giant ratio. We vary this ratio by changing the slope ($\Gamma_\mathrm{b}$) of the high-mass end of the IMF, while it was kept constant for stars with masses below $0.5 M_\sun$. Notice that our results, restricted to inference of the dwarf-to-giant ratio, barely depend on the detailed IMF shape, but it must be considered when exploring, for example, the chemical evolution and the expected mass-to-light ratios of ETGs.

\subsection{The underlying parameters behind the varying dwarf-to-giant ratio in ETGs}
Two competing candidates have been proposed to explain the observed dwarf-to-giant ratio variations in ETGs: $\sigma$ and [Mg/Fe]. \citet{Smith} investigated these two parameters over the same sample of galaxies \citep{atlas3d,cvd12}, pointing out that the stellar population analysis favors the stellar population property ([Mg/Fe]), whereas a dynamical analysis supports the dynamical-related quantity ($\sigma$). The work of \citeauthor{Smith} opened the question of whether IMF studies are actually probing the IMF or whether they are, at least partially, dominated by confounding factors. However, stellar population analyses of SDSS stacked spectra \citep{labarbera,spiniello2013} have shown a strong correlation between $\sigma$ and the dwarf-to-giant ratio. Moreover, \citet{labarbera15}, using also SDSS stacked spectra, showed that [Mg/Fe] is loosely (or un-) correlated with the IMF slope. The latter result is in agreement with panel (c) in Fig.~\ref{fig:all}, where we show that the local [Mg/Fe] is decoupled from the dwarf-to-giant ratio variations. 

\begin{figure}
\begin{center}
\includegraphics[width=8.5cm]{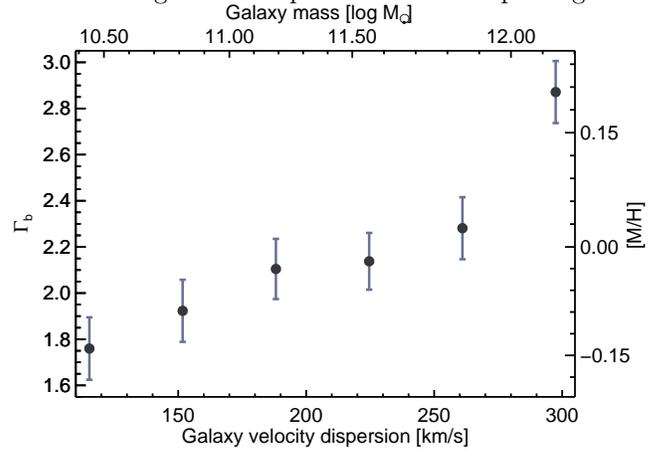}
\end{center}
\caption{Predicted IMF slope -- galaxy velocity dispersion relation. The best-fitting metallicity of the global SDSS measurements (right vertical axis) was transformed, following Eq.~\ref{eq:fit}, into an {\it expected} IMF-slope value $\Gamma_\mathrm{b}$. The upper horizontal axis represents the galaxy mass, estimated from the velocity dispersion \citep{Cappellari13}. We suggest that the relation found between IMF slope and galaxy mass can be understood as the combination of the mass -- metallicity and the metallicity -- IMF slope relations.}
\label{fig:relation}
\end{figure}

Regarding $\sigma$, in \citet{mn14b} we showed that its local value was not the main driver behind the dwarf-to-giant ratio variations of the massive relic galaxy NGC~1277. In the present work we confirm this result using a statistically significant sample of ETGs (panel (a) in Fig.~\ref{fig:all}). Notice that previous studies suggesting a connection between $\sigma$ and the IMF slope were based on spatially unresolved spectra. Thus, their velocity dispersions trace the overall galaxy potential (mass) rather than the detailed kinematics. Thanks to the CALIFA data, we propose that this connection between galaxy mass and IMF slope arises from considering simultaneously both the galaxy mass--metallicity and the metallicity--IMF slope relations. In this sense, the dwarf-to-giant ratio gradients observed in ETGs \citep{mn14,mn14b} can be also accounted for the radial metallicity variation within these galaxies. To illustrate the suggested metallicity-driven galaxy mass--IMF slope relation, we used the above described Sloan data (black dots in Fig.~\ref{fig:met}) to calculate the mean global metallicity per galaxy mass ($\sigma$) bin. Then, we transformed this metallicity into the expected IMF slope according to Eq.~\ref{eq:fit}. The result is summarized in Fig.~\ref{fig:relation}. Although it can not be directly compared with previous works, we qualitatively recover the observed relation between galaxy mass and IMF slope, with more massive galaxies being bottom-heavier.

\subsection{Metallicity as a driver of IMF variations}
Metallicity-driven dwarf-to-giant ratio variations can be explained if molecular clouds collapse following a Jeans spectrum. Such a scenario is expected if thermal pressure dominates the process, although a turbulent medium can yield similar results if dust cooling is significant \citep{larson:05}. Perhaps, a realistic scenario would couple a Jeans-driven mechanism with turbulence-induced fragmentation \citep[e.g.][]{Padoan}. Nevertheless, this letter shows the importance that metallicity plays in determining the stellar mass spectrum. Such a scenario is consistent with studies of nearby resolved galaxies \citep{Geha}, line-strength indices \citep{cenarro} and optical colors \citep{Ricciardelli}. In low-metallicity regions, it is found that molecular clouds are very efficient in forming massive stars, leading to top-heavy IMFs. This is supported by observations of metal-poor systems showing evidence of an IMF dominated by massive stars \citep{Marks}.

If metallicity is one of the main drivers behind the IMF variations, a number of fundamental aspects of galaxy evolution must be revisited. Because the stellar feedback regulates the chemical enrichment of galaxies, IMF, star formation and chemical composition will be tightly related during the time evolution. In particular, massive galaxies are expected to increase their metallicity during the formation of their stellar populations, and therefore, the dwarf-to-giant ratio should have been smaller at higher redshifts, such that there were fewer low-mass stars per high-mass star. Such a time-dependent IMF scenario \citep{Arnaud,vazdekis:97,Larson,weidner:13,Ferreras15} would explain the observations of nearby massive ETGs, which are inconsistent with a time-invariant, steep (or even Milky-Way like) IMF slope \citep{Arrigoni}.

Moreover, the [Mg/Fe] has been extensively used as a proxy for the formation time-scale \citep[e.g.][]{thomas05}: rapid formation events lead to high [Mg/Fe] values. However, an IMF dominated at early times by high-mass stars would also produce an enhanced [Mg/Fe], relaxing the typical ($\sim1$Gyr) constraint on the rapid formation time-scale. Thus, to safely interpret $z\sim0$ observations, it is crucial to understand the chemical evolution of galaxies since they were formed, and in particular, during the build-up of their chemical properties \citep{vazdekis96}. In this sense, it has been shown \citep[e.g.][]{Ferreras09,choi,anneta} that the chemical composition of massive galaxies has remained constant over the last $\sim7$ Gyr. If metallicity actually regulates the dwarf-to-giant ratio variations, it would imply that the IMF of massive objects was bottom-heavy at $z\sim1$, as recently suggested by observations \citep{Shetty,z1}, since no chemical evolution has happened since then.

Finally, we want to emphasize that the present study is restricted to the analysis of ETGs. The local conditions (e.g. turbulence, pressure, density, radiation, and magnetic fields) of the interstellar medium at high-$z$, or even in nearby disk galaxies, are expected to differ from the local environment within ETGs. Thus, more observational efforts are still needed to fully characterize the IMF behavior with varying star-forming conditions. In fact, for very metal-poor systems, \citet{Geha} found a slightly top-heavy IMF, but steeper than expected from a crude extrapolation of Eq.~\ref{eq:fit}. The trend in Fig.~\ref{fig:met} seems to actually flatten at lowest and highest metallicity, pointing to a non-linearity of  the IMF--metallicity relation.

\section{Summary}

Although deviations from the standard Milky-Way IMF have been extensively reported in the literature over the last years, the mechanism responsible for these variations remains unknown. We have analysed the radial gradients of IMF-sensitive features in a sample of 24 ETGs observed by the CALIFA survey, finding that {\it the local IMF is tightly related to the local metallicity.} Our result agrees with previous works reporting local and global IMF variations, and it explains the observed galaxy mass--IMF relation. We speculate about the implications of metallicity-driven IMF variations in the context of galaxy formation and evolution.

The intimate connection between IMF and metallicity described in our work suggests a complex massive galaxy formation process, departing from the classical picture where these objects formed under nearly stationary conditions. Thus, to safely understand and interpret $z\sim0$ observations, it is necessary to untangle the early life of massive galaxies at high-$z$. The combination of more sophisticated stellar population synthesis models \citep{cvd12,alpha} and high-$z$ 
spectroscopic surveys \citep{Brammer,mosdef} will provide valuable insights. \looseness-2

\acknowledgments 
We acknowledge support from grants AYA2013-48226-C3-1-P and AYA2010-15081 from the Spanish MINECO. CJW acknowledges support through the Marie Curie Career Integration Grant 303912. P.S-B acknowledges support from the Ram\'on y Cajal program, ATA2010-21322-C03-02 (MINECO). JFB and GvdV acknowledge DAGAL network from the People Programme (Marie Curie Actions) of the European Unions Seventh Framework Programme FP7/2007-2013 under REA grant agreement number PITN-GA-2011-289313. We thank the referee for his/her constructive review.

% \bibliographystyle{apj}
% \bibliography{califa}

\end{document}